\documentclass[aps,prl,twocolumn,superscriptaddress,floatfix]{revtex4-1}

\usepackage{graphicx}
\usepackage{dcolumn}
\usepackage{bm}

\usepackage[T1]{fontenc}
\usepackage{amsmath}
\usepackage{amssymb}
\usepackage{graphicx}

\usepackage[utf8]{inputenc}
\usepackage[T1]{fontenc}
\usepackage{mathptmx}
\usepackage{etoolbox}

\makeatletter

\providecommand{\tabularnewline}{\\}

\makeatother

\usepackage{babel}
\begin{document}
\title{A planning tool for neutron powder diffraction experiments}
\author{Joseph A. M. Paddison}
\email{paddisonja@ornl.gov}
\affiliation{Neutron Scattering Division, Oak Ridge National Laboratory, Oak Ridge, TN 37831, USA}%
\author{Stuart Calder}
\affiliation{Neutron Scattering Division, Oak Ridge National Laboratory, Oak Ridge, TN 37831, USA}%
\author{Danielle R. Yahne}
\affiliation{Neutron Scattering Division, Oak Ridge National Laboratory, Oak Ridge, TN 37831, USA}%
\author{Malcolm J. Cochran}
\affiliation{Neutron Scattering Division, Oak Ridge National Laboratory, Oak Ridge, TN 37831, USA}%
\author{Si Athena Chen}
\affiliation{Neutron Scattering Division, Oak Ridge National Laboratory, Oak Ridge, TN 37831, USA}%
\author{Matthias D. Frontzek}
\affiliation{Neutron Scattering Division, Oak Ridge National Laboratory, Oak Ridge, TN 37831, USA}%
\author{Yuanpeng Zhang}
\email{zhangy3@ornl.gov}
\affiliation{Neutron Scattering Division, Oak Ridge National Laboratory, Oak Ridge, TN 37831, USA}%

\begin{abstract}
We introduce a computer program to simulate the results of neutron
powder-diffraction experiments at the High Flux Isotope Reactor at
Oak Ridge National Laboratory. The program is freely available as
a web application at \emph{http://addie.ornl.gov/hfirestimate, }and
is designed to be straightforward to use for researchers who are new
to neutron diffraction. The input includes the crystal structure of
the proposed sample, the sample mass, and the instrument configuration.
The results include a plot of the simulated data -- including realistic
estimates of background and the error bars due to counting statistics
-- and suggestions of how to resolve potential problems with the
experiment. Here, we explain the design and implementation of this
program and demonstrate its performance using comparisons of simulated
and experimental data. We hope that this program will enable researchers
to plan neutron-scattering experiments more effectively, increasing
the likelihood of successful experiments and improving the productivity
of neutron-diffraction research.
\end{abstract}
\maketitle

\section{Introduction}

Neutron powder diffraction is a pre-eminent technique for the refinement
of crystal and magnetic structures of materials, due to the high sensitivity
it offers to the positions of light elements and the arrangements
of magnetic moments \citep{Boothroyd_2020,Squires_1978}. However,
neutron diffractometers require either a nuclear reactor or an accelerator-driven
spallation source to generate neutrons, and so are expensive to build
and operate. For these reasons, neutron scattering instruments are
typically located at central facilities or national laboratories,
and access to these facilities is obtained through a competitive peer-review
process with only a few opportunities per year to submit proposals.
Consequently, neutron-scattering researchers -- those who use neutron
facilities and those who operate them -- share the goal of ensuring
that effective use is made of the limited available beam-time.

To help realise this goal, substantial progress has been made in optimising
neutron instrumentation \citep{Lefmann_1999,Willendrup_2020} and
data analysis \citep{Arnold_2014,Rodriguez-Carvajal_1993a,Toby_2013}.
However, less attention has been given to what is arguably an equally
important factor -- the \emph{planning} of neutron-diffraction experiments.
Planning is essential to the success of any experiment, and there
are several ways in which insufficient planning can cause a neutron-diffraction
experiment to fail to produce analysable data. Neutron diffraction
measurements are often limited by neutron flux, and samples that are
too small can lead to an inability to detect a signal of interest.
Conversely, some isotopes are strong neutron absorbers, and failure
to optimise the sample size or shape in such cases can lead to negligible
neutron scattering due to strong absorption. Some elements, such as
hydrogen and vanadium, have large incoherent neutron-scattering cross
sections that generate a large effective background and obscure the
structural information.

At present, experiment planning to avoid these pitfalls typically
involves a conversation between the visiting researcher and the facility
scientist who is responsible for the instrument. The instrument scientist's
guidance is typically qualitative and informed by their experience
of measuring similar samples previously. In marginal or
novel cases, however, the likelihood of success for the experiment
may be difficult to predict, since existing tools to simulate Bragg
scattering \citep{Rodriguez-Carvajal_1993a,Toby_2013} do not account
for the count rate or background of specific instruments. Moreover,
if the visiting scientist does not have experience using the instrument,
they may be surprised (whether favourably or not) by the quality of
the data they actually measure.

To address these problems, we developed a software tool for planning
experiments using the neutron powder diffractometers at the High Flux
Isotope Reactor (HFIR) at ORNL. The design criteria were that the
software should produce an accurate prediction of the powder diffraction
pattern -- including statistical uncertainties and background counts
-- given minimal input from the user. The main input would include
the crystal structure and mass of the sample to be measured; the planned
measurement time; and the instrument configuration to be used. The
calculation should also take no more than a few seconds of CPU time,
which would allow it to be provided as a small web application. 

We successfully developed an experiment planning tool that meets these
criteria, which is available at \emph{}\\
\emph{http://addie.ornl.gov/hfirestimate}. In this paper, we discuss
the principles and implementation of the tool. While the principles
are already well understood, they have previously been applied to
correction of measured data \citep{Soper_2011} rather than to simulation
of potential future experiments. We start by discussing the possible
approaches to simulating powder diffraction data and introduce the
interface and required input. We explain how each contribution to
the scattering is calculated, and benchmark the results by comparing
simulated and experimental data for Si and Ba$_{4}$CeMn$_{3}$O$_{12}$
(previously published in Ref.~\citep{Regier_2024}). Finally, we
explore possible applications and extensions of this approach.

\section{General Approach}

\begin{figure*}[t]
\includegraphics[width=16cm]{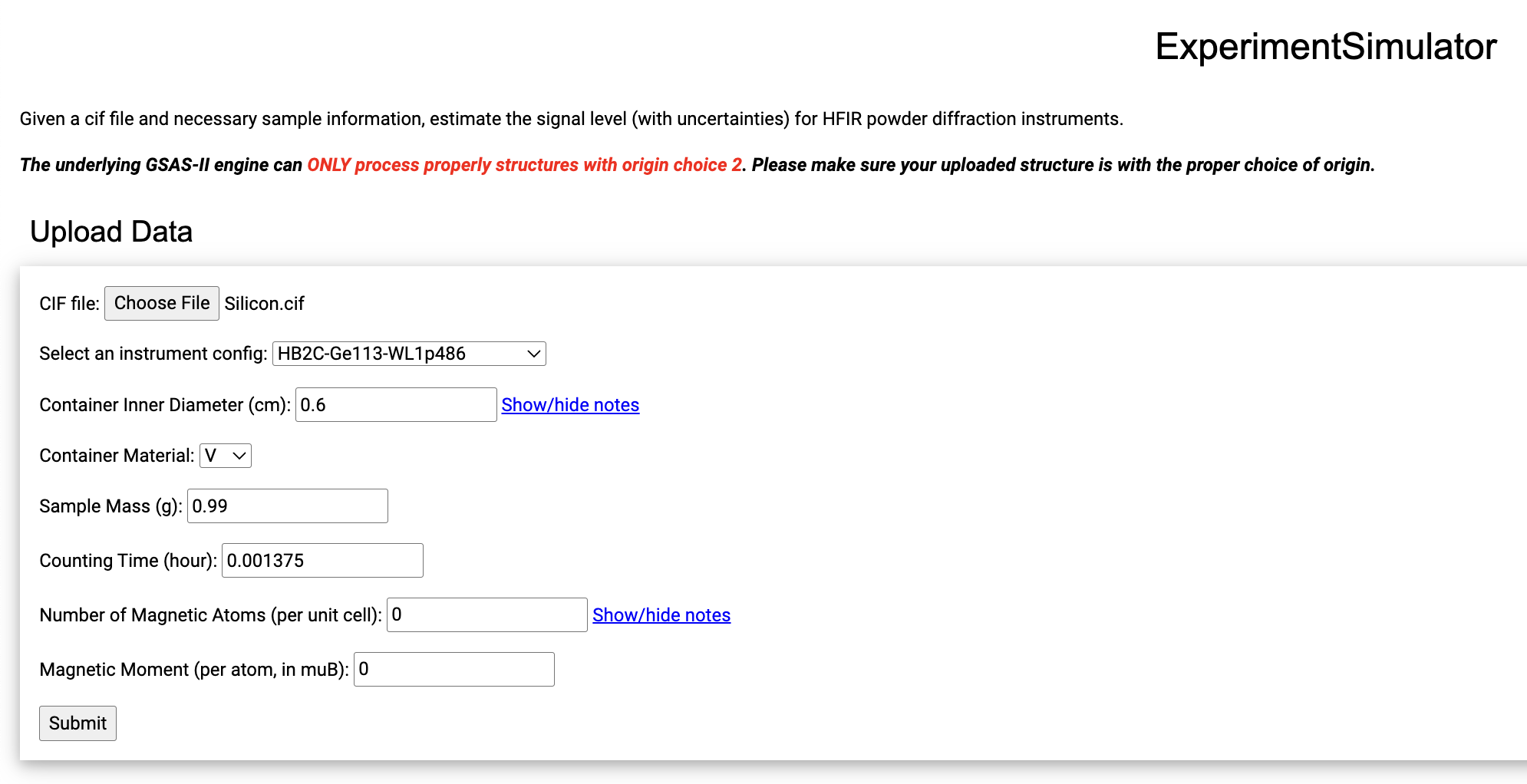}

\caption{Start page for web application, showing the necessary input. The input
parameters were used to calculate the Si example discussed in Section~\ref{sec:Output-and-Examples}.}

\end{figure*}

For simplicity, we consider only the constant-wavelength neutron diffraction
instruments located at the High Flux Isotope Reactor (HFIR) at Oak
Ridge National Laboratory \citep{Calder_2018}. These are HB-2A (POWDER)
\citep{Garlea_2010} and HB-2C (WAND$^{2}$) \citep{Frontzek_2018,Frontzek_2018a}.
The HB-2A instrument is a powder diffractometer that is often used
to study magnetic structures. It has three commonly-used incident
wavelengths, $2.41$, $1.54$, or $1.12$~$\text{Å\ }$, each of
which may be operated with or without a pre-sample collimator. With
$1.54$-$\text{Å\ }$ neutrons the resolution is $\sim$$0.38^{\circ}$
FWHM at the Si $(220)$ peak. The current detector array contains
gaps and hence must be moved to collect a full powder diffraction
pattern, which is typically measured over the range $5^{\circ}<2\theta<127^{\circ}$.
A planned detector upgrade, termed MIDAS, is aimed at increasing data
collection rates by an order of magnitude. The simulator will be appropriately
updated after the upgrade. The HB-2C instrument is a high count rate
powder diffractometer with single-crystal capabilities. It typically
uses an incident wavelength of $1.486$\,$\text{Å}$, giving a resolution
of $\sim$$0.63^{\circ}$ FWHM at the Si $(220)$ peak. It has continuous
detector coverage, and the detector is typically located to measure
over the range $5^{\circ}<2\theta<125^{\circ}$ .

We considered two potential approaches for simulating such experiments.
One possibility is to build a ``virtual instrument'' in a program
such as McStas \citep{Lefmann_1999,Willendrup_2020} or McVine \citep{Lin_2016},
which models each physical component and estimates the scattering
using a Monte Carlo ray-tracing approach. This method is widely used
for simulating the performance of new instrument designs, but is computationally
expensive. The second possibility is to develop a phenomenological
model of the existing HFIR instruments using calibration measurements.
We chose this latter approach since it allows for faster simulations.

\section{Calibration}

We used three calibration measurements to characterise each instrument's
performance---a vanadium standard, an empty instrument measurement,
and a Si powder standard. The results of the instrument calibration
are summarised in Table \ref{tab:calibration}.

\begin{table*}

\begin{tabular}{c|ccccc}
\hline 
Instrument configuration & Count rate (counts/h per b/sr) & $U$ & $V$ & $W$ & $X$\tabularnewline
\hline 
\hline 
\textbf{HB-2A, $\lambda=2.41$\,$\text{Å\ }$, collimator out} & $1.11\times10^{-18}$ & 0.68 & $-0.61$ & $0.26$ & $0.063$\tabularnewline
HB-2A, $\lambda=1.54$\,$\text{Å\ }$, collimator out & $2.71\times10^{-18}$ & $0.61$ & $-0.60$ & $0.26$ & $0.10$\tabularnewline
HB-2A, $\lambda=1.12$\,$\text{Å\ }$, collimator out & $1.36\times10^{-18}$ & $0.54$ & $-0.56$ & $0.24$ & $0.14$\tabularnewline
\textbf{HB-2C, $\lambda=1.486$\,$\text{Å}$} & $7.25\times10^{-18}$ & $2.96$ & $-1.94$ & $0.75$ & $0$\tabularnewline
\hline 
\end{tabular}\caption{\label{tab:calibration}Instrument properties estimated from calibration
measurements. Configurations shown in \textbf{bold }are most frequently
used. For HB-2A, count rate is given per detector. For HB-2C, count
rate is given per $0.1^{\circ}$. For HB-2A, the count rate with the
pre-sample collimator in is approximately 42\% of the values given
for the collimator out. }
\end{table*}

\subsubsection{Vanadium standard}

Since V has a large incoherent cross-section and a small coherent
cross-section, it can be used to estimate the count rate of the instrument.
For a given detector and instrument configuration, we determine the count rate as
\begin{equation}
C=\frac{V-B}{(N_{\mathrm{V}}/4\pi)[\sigma_{\mathrm{V}}T+\sigma_{\mathrm{mult}}]},\label{eq:count_rate}
\end{equation}
 where $V$ denotes counts per hour on the vanadium standard sample;
$B$ denotes background counts/h as obtained from the empty instrument
measurement; $N_{\mathrm{V}}$ denotes the number of V atoms in the
sample; $\sigma_{\mathrm{V}}$ denotes the incoherent scattering cross
section in barn (b) per V atom; $T$ denotes the (angle-dependent)
transmission; and $\sigma_{\mathrm{mult}}$ denotes the calculated
multiple scattering cross section. The calculation of the transmission
and multiple scattering is discussed below. The numerator of Eq.~(\ref{eq:count_rate})
is the measured count rate on the vanadium sample, corrected for background.
The denominator is the calculated cross-section in b per steradian.
Consequently, for a given instrument configuration, Eq.~(\ref{eq:count_rate})
gives the count rate per unit scattering cross-section, expressed
as counts/detector/h per b/sr. This is the key quantity for simulations,
since it can be used to estimate the counting statistics for \emph{any}
scattering signal that can be calculated in absolute intensity units.

\subsubsection{Empty instrument measurement}

Empty instrument measurements were used to characterise the intrinsic
background for each instrument configuration. A reasonable description
of the measured background was given by the function
\begin{equation}
B(2\theta)=a_{0}+b_{0}(2\theta-2\theta_{0})^{2}+c_{0}(2\theta-2\theta_{0})^{4},\label{eq:fit_bg}
\end{equation}
where $a_0$, $b_0$, $c_0$ and $2\theta_{0}$ are fit parameters. 

\subsubsection{Silicon standard}

Measurements of a NIST silicon standard were used to characterise
the resolution function (i.e., the peak shape) for each instrument
configuration. This was done by performing a Rietveld refinement to
the Si standard data using the FullProf program \citep{Rodriguez-Carvajal_1993a},
and refining the pseudo-Voigt peak-shape parameters $U$, $V$, $W$,
and $X$ defined in Ref.~\citep{Rodriguez-Carvajal_2001}.

\section{Input and Interface }

The input to the simulation is the following:
\begin{itemize}
\item The instrument configuration, including the instrument (HB-2A or HB-2C),
and the neutron wavelength $\lambda$. 
\item The crystal structure in .cif format. This is parsed to obtain the
absorption, incoherent, and total scattering cross sections, $\sigma_{\mathrm{abs}}$,
$\sigma_{\mathrm{inc}}$, and $\sigma_{\mathrm{sc}}$ respectively,
in b per unit cell. We note that the absorption cross-section is dependent
on $\lambda$. The cif is also used to obtain the crystallographic
density $\rho$. It is assumed that isotopes are present in their
natural abundance, except in the case of hydrogen, which can be specified
as H (for $^{1}$H) or D (for $^{2}$H) in the .cif.
\item The sample mass in g. This is used to estimate the total number of
unit cells in the sample, $N_{\mathrm{s}}$, and the number of unit
cells per unit volume, $n_{\mathrm{s}}$. This calculation assumes
a packed density of $0.5\rho$, as is typical for powder samples.
\item The sample container material, which can be V, Al, or Cu.
\item The inner diameter $2r$ of the sample container in cm. The container
is assumed to be a cylinder and typical diameters are 0.25 to 1.0\,cm. For the standard containers used at HB-2A and HB-2C, the thickness
of the container wall is 0.0381\,cm for Al and Cu containers, and
approximately 0.015 cm for V containers. This information is used
(with the sample mass and packed density) to estimate the sample height
$h$, and the number of unit cells $N_{\mathrm{can}}$ of container
material in the beam.
\item The measurement duration in h, notated $t_{\mathrm{meas}}$.
\item For magnetic systems, the number of magnetic atoms per unit cell.
If magnetic atoms fully occupy a single crystallographic site, this
is the multiplicity of that site.
\item For magnetic systems, the magnetic moment per magnetic atom, in $\mu_{\mathrm{B}}$.
\end{itemize}
The frontend interface implemented here is part of the web-based ADvanced
DIffraction Environment (ADDIE), available at \emph{https://addie.ornl.gov}.
It uses the Python Flask framework as the backend, and aims to provide
neutron and X-ray scattering users with an easy-access platform for
experimental planning, data processing, and some straightforward data-analysis
tasks. Specifically for the experiment simulator tool discussed here,
two auxiliary programs were used in the workflow. The GSAS-II scriptable
interface \citep{Toby_2013} was used for the powder diffraction pattern
simulation. In GSAS-II, the PyCifRW module \citep{Hester_2006} is
used as the CIF parser, which can sometimes fail in processing CIF
files that do not follow the required standard. To cope with such
exceptions, the headless VESTA interface \citep{Momma_2011} was used
as the engine for the initial CIF parsing: the user uploaded CIF file
is imported into VESTA and exported to a VESTA version of the CIF,
which is then readable by PyCifRW (and therefore GSAS-II).

\section{Simulation}

We assume that the data can be simulated as the sum of contributions
from sample and background (including sample container):
\begin{align}
I^{\mathrm{sim}} & =I_{\mathrm{s}}^{\mathrm{sim}}+I_{\mathrm{b}}^{\mathrm{sim}}.
\end{align}
This assumption is reasonable provided the sample is not strongly
absorbing. If the sample is absorbing, the simulated background level
is overestimated, since the simulator does not account for attenuation
of the background by the sample.

\subsection{Sample contributions to the scattering}

Throughout, we denote intensities in experimental
units (counts/h) with superscript ``sim'', and intensities in absolute units (b/sr per unit
cell) without a subscript. For scattering from the
sample, these quantities are related as
\begin{equation}
I_{\mathrm{s}}^{\mathrm{sim}}=t_{\mathrm{meas}}N_{\mathrm{s}}CI_{\mathrm{s}},
\end{equation}
where $t_{\mathrm{meas}}$ is the measurement duration in h, $C$ is
the count rate in the detector given by Eq.~(\ref{eq:count_rate}),
$I_{\mathrm{s}}$ is the absolute scattering intensity in b/sr per
unit cell, and we recall that $N_{\mathrm{s}}$ is the number of unit
cells in the sample. In absolute units, the scattering signal from
the sample is given by
\begin{equation}
I_{\mathrm{s}}=I_{\mathrm{inc}}+I_{\mathrm{nuc}}+I_{\mathrm{mag}}+I_{\mathrm{mult}},
\end{equation}
where $I_{\mathrm{nuc}}$ is the nuclear Bragg intensity, $I_{\mathrm{inc}}$
is the incoherent scattering intensity, $I_{\mathrm{mag}}$ is the
magnetic scattering intensity, and $I_{\mathrm{mult}}$ is the multiple
scattering intensity. We now consider each of these terms in turn.

\subsubsection{Incoherent scattering}

The incoherent scattering intensity from the sample is given by
\begin{equation}
I_{\mathrm{inc}}(2\theta)=T(2\theta)\frac{\sigma_{\mathrm{inc}}}{4\pi},
\end{equation}
where $\sigma_{\mathrm{inc}}$ is the incoherent scattering cross-section
in b/sr per unit cell, obtained from tabulated values \citep{NN_1992}.
For example, for Si with 8 atoms in its unit cell, $\sigma_{\mathrm{inc}}=8\times0.004=0.032$
b/sr per unit cell. 

The transmission $T$ is a function of the dimensionless product $\mu r$,
where $\mu=n_{\mathrm{s}}(\sigma_{\mathrm{abs}}+\sigma_{\mathrm{sc}})$
is the linear attenuation coefficient and $r$ is the container's
inner radius. For samples with $\mu r\lesssim1$, a simple formula
given by Hewat \citep{Hewat_1979} can be used to calculate $T(2\theta)$.
For more strongly absorbing samples, a more general formula given
by Sabine and Hunter is used \citep{Sabine_1998}.

\subsubsection{Nuclear Bragg scattering}

The nuclear Bragg scattering can be calculated, in b/sr per unit cell,
as
\begin{equation}
I_{\mathrm{nuc}}(2\theta)=T(2\theta)\frac{45\lambda^{3}}{4\pi^{2}V_{0}}\sum_{\mathbf{G}}\frac{m_{\mathbf{G}}\left|F_{\mathbf{G}}\right|^{2}}{\sin^{2}\theta_{\mathbf{G}}\cos\theta_{\mathbf{G}}}R(2\theta-2\theta_{\mathbf{G}}),\label{eq:abs_2th}
\end{equation}
where $F_{\mathbf{G}}$ is the unit-cell structure factor, $V_{0}$
is the volume of the unit cell, $\mathbf{G}$ labels a set of equivalent
Bragg reflections having multiplicity $m_{\mathbf{G}}$ and scattering
angle $2\theta_{\mathbf{G}}$, and $R(2\theta-2\theta_{\mathbf{G}})$
is the resolution function that describes the pseudo-Voigt peak shape
\citep{Boothroyd_2020,Paddison_2025}. Rather than calculating Eq.~(\ref{eq:abs_2th})
directly, we used the scripting interface to the Rietveld-refinement
program GSAS-II \citep{Toby_2013} to perform this calculation using
the input .cif and peak-shape parameters (see Table~\ref{tab:calibration}).
However, since GSAS-II does not calculate the nuclear Bragg scattering
in b/sr per unit cell, it was necessary to convert the GSAS-II calculation
result, $I_{\mathrm{nuc}}^{\mathrm{GSAS}}(2\theta)$. The conversion
factor is derived in Ref.~\citep{Paddison_2025} and is given by
\begin{equation}
I_{\mathrm{nuc}}(2\theta)=T(2\theta)\frac{4500\lambda^{3}}{2\pi^{2}}I_{\mathrm{nuc}}^{\mathrm{GSAS}}(2\theta).
\end{equation}

\subsubsection{Magnetic scattering}

The goal of powder-diffraction experiments on magnetic materials is
often to determine the magnetic structure. It is important that the
planning tool can predict the strength of the magnetic signal in such
measurements. Since the magnetic structure is typically not known
before the experiment, we calculate the magnetic scattering for two
simple cases: a paramagnet and a ferromagnet. In these cases, the
only information that is needed is the number of magnetic atoms in
the crystallographic unit cell, $N_{\mathrm{mag}}$, and the magnetic moment
per atom (in $\mu_{\mathrm{B}}$). The former is usually apparent
from the crystal structure, and the latter can be estimated from magnetometry
data; e.g., Curie-Weiss fits to the bulk magnetic susceptibility yield
an estimate of the squared effective magnetic moment given by
\begin{equation}
\mu_{\mathrm{eff}}^{2}=\mu_{\mathrm{ord}}(\mu_{\mathrm{ord}}+2),\label{eq:mu_eff}
\end{equation}
 where $\mu_{\mathrm{ord}}$ is the ordered magnetic moment that should
be input for simulations of magnetic Bragg scattering.

The paramagnetic scattering is given, in b/sr per unit cell, by
\begin{equation}
I_{\mathrm{paramag}}(2\theta)=T(2\theta)\frac{2}{3}\left(\frac{\gamma_{n}r_{e}}{2}\right)^{2}N_{\mathrm{mag}}\mu_{\mathrm{eff}}^{2}f_{\mathrm{mag}}^{2}(2\theta),\label{eq:paramaga}
\end{equation}
where the constant $\left(\gamma_{n}r_{e}/2\right)^{2}=0.07265$ b
\citep{Squires_1978,Boothroyd_2020}. The squared magnetic form factor
$f_{\mathrm{mag}}^{2}(2\theta)$ depends on the magnetic ion, but
for simplicity, we take $f_{\mathrm{mag}}^{2}(2\theta)\sim\exp(-0.05Q^{2})$
as typical. We can therefore define a magnetic scattering cross-section
\begin{equation}
\sigma_{\mathrm{mag}}=\frac{8\pi}{3}\left(\frac{\gamma_{n}r_{e}}{2}\right)^{2}N_{\mathrm{mag}}\mu_{\mathrm{ord}}^{2},
\end{equation}
in b per unit cell, which can be compared to the nuclear coherent
scattering cross-section, $\sigma_{\mathrm{sc}}-\sigma_{\mathrm{inc}}$. 

The ferromagnetic scattering is approximated by a simple scaling of
the nuclear Bragg scattering,
\begin{equation}
I_{\mathrm{ferromag}}(2\theta)=\frac{\sigma_{\mathrm{mag}}}{\sigma_{\mathrm{sc}}-\sigma_{\mathrm{inc}}}f_{\mathrm{mag}}^{2}(2\theta)I_{\mathrm{nuc}}^{\mathrm{abs}}(2\theta).
\end{equation}
While this calculation typically does not correspond to a physically
meaningful magnetic structure, it gives a good indication of the magnitude
of the magnetic scattering magnitude relative to the nuclear Bragg
intensities, which is most important for experiment planning.

\subsubsection{Multiple scattering}

The multiple scattering cross section is given by
\begin{equation}
\sigma_{\mathrm{mult}}=\frac{\sigma_{\mathrm{sc}}(\sigma_{\mathrm{sc}}/\sigma_{\mathrm{tot}})\delta}{1-(\sigma_{\mathrm{sc}}/\sigma_{\mathrm{tot}})\delta},
\end{equation}
where $\delta$ is a function of $\mu r$ and the radius-to-height
ratio $r/h$ \citep{Blech_1965}, and $\sigma_{\mathrm{tot}}=\sigma_{\mathrm{sc}}+\sigma_{\mathrm{abs}}$.
To obtain a rapid estimate of $\delta$, we interpolate the tabulated
values given by Blech and Averbach \citep{Blech_1965}.

\subsection{Background contributions to the scattering}

\begin{figure}[t]
\includegraphics{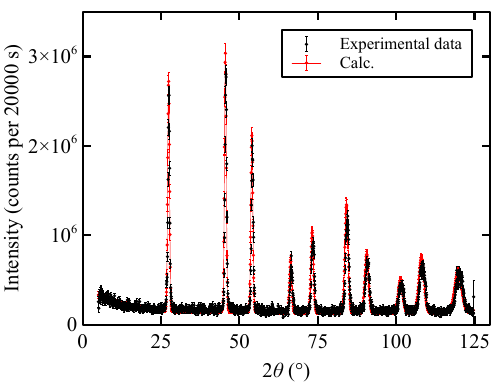}

\caption{\label{fig:si_sim}Comparison of experimental HB-2C data (black points)
and simulation (red points and line) for a Si standard sample. The
estimated sample mass was 0.99\,g, the measurement time was 5\,s,
and the instrument used an incident neutron wavelength of 1.486\,$\text{Å}$. }
\end{figure}

\begin{figure*}[t]
\includegraphics{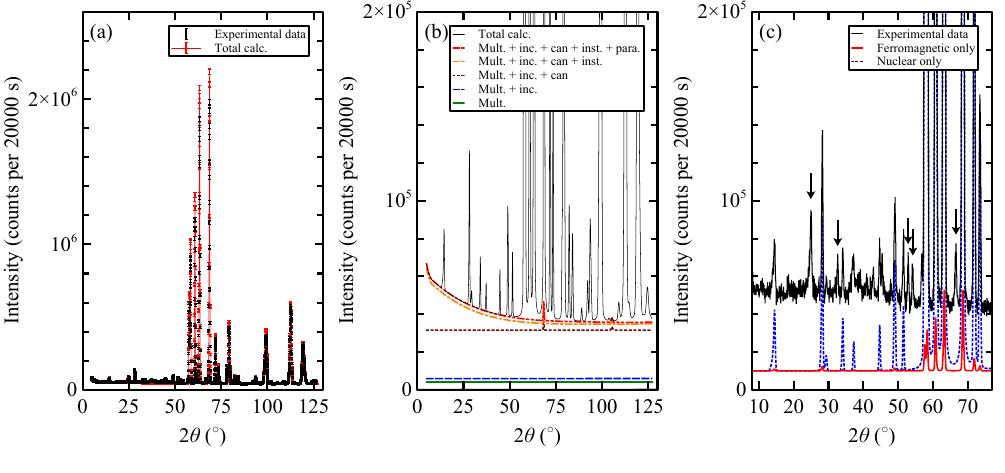}

\caption{\label{fig:bcm_sim}(a) Comparison of experimental HB-2A data measured
on Ba$_{4}$CeMn$_{3}$O$_{12}$ at $T\approx4$\,K with simulation.
The data were previously published in \citep{Regier_2024}. For the
measurement and the simulation, the sample mass was 4.2~g, the measurement
time was 1.87\,h, and the instrument configuration used an incident
neutron wavelength of $2.41$\,$\text{Å}$ with pre-sample collimator
out. (b) Simulated pattern (same as in (a)), shown on an expanded
vertical scale to highlight the different contributions to the apparent
background. Labels are as follows: ``mult.'' = multiple scattering,
``inc.'' = incoherent scattering, ``can'' = background from sample
container, ``inst.'' = background from instrument, and ``para.''
= paramagnetic scattering assuming a magnetic moment of $1.1\,\mu_{\mathrm{B}}$
per Mn. Colours and line styles are labelled in the plot. (c) Comparison
of simulated nuclear intensity (blue dotted line), simulated ferromagnetic
intensity assuming a magnetic moment of $1.1\,\mu_{\mathrm{B}}$ per
Mn (red solid line), and experimental data (black solid line).}
\end{figure*}

The background scattering intensity is estimated as
\begin{equation}
I_{\mathrm{b}}^{\mathrm{sim}}=I_{\mathrm{inst}}^{\mathrm{sim}}+I_{\mathrm{can}}^{\mathrm{sim}},\label{eq:bg}
\end{equation}
where $I_{\mathrm{inst}}^{\mathrm{sim}}$ is the instrument background
and $I_{\mathrm{can}}^{\mathrm{sim}}$ is the background from
the sample container. It would be possible to include additional contributions
(e.g. from a cryostat or furnace), but this is not currently done.
Since the sample environment attenuates the beam and adds background,
Eq.~(\ref{eq:bg}) gives the best-case background for a non-absorbing
sample. We note, however, that both HB-2A and HB-2C have collimation
that reduces the background from the sample environment.

\subsubsection{Instrument background}

The empty-instrument background is obtained from Eq.~(\ref{eq:fit_bg})
by the relation
\begin{equation}
I_{\mathrm{inst}}^{\mathrm{sim}}(2\theta)=\frac{2rht_{\mathrm{meas}}}{A_{0}}B(2\theta),
\end{equation}
where $A_{\mathrm{0}}$ is the beam area for the calibration background
measurement, and the beam area for the sample measurement is approximated
by the area of the sample, $2rh$. This approximation is
motivated by the fact that, in a typical measurement, the beam area
is defined by adjustable slits that are set to match the sample size.

\subsubsection{Sample container background}

The sample container background is calculated in the same way as the
scattering from the sample; i.e., as the sum of nuclear Bragg and
incoherent contributions from the container material (either V, Cu,
or Al). The simulated intensity is given by
\begin{equation}
I_{\mathrm{can}}^{\mathrm{sim}}(2\theta)=t_{\mathrm{meas}}N_{\mathrm{can}}CI_{\mathrm{can}}(2\theta),
\end{equation}
 where $N_{\mathrm{can}}$ is the number of unit cells of container
material.

\section{\label{sec:Output-and-Examples}Output and Examples}

\subsection{Example 1: Nuclear scattering from Si on HB-2C}

To benchmark the simulator's performance, we compare simulated diffraction
patterns with experimental data measured on a polycrystalline Si standard.
The data were measured for 5\,s using an incident wavelength of 1.486\,$\text{Å}$
on the HB-2C instrument. The sample was loaded in a vanadium container
of 0.6-cm diameter, and the vertical slits were set to illuminate
a sample height of 3\,cm. Assuming a 50\% packing fraction, we estimate
that 0.99\,g of Si powder was illuminated by the neutron beam. The
simulated neutron-diffraction pattern is compared with the HB-2C data
in Figure~\ref{fig:si_sim}. There is excellent overall agreement
between the simulation and the experimental data, both in terms of
the statistical quality (i.e., the magnitudes of the error bars) and
the overall background level. 

\subsection{Example 2: Nuclear and magnetic scattering from Ba$_{4}$CeMn$_{3}$O$_{12}$
on HB-2A}

As a second test, we compare simulated diffraction patterns with experimental
HB-2A data measured on the layered perovskite Ba$_{4}$CeMn$_{3}$O$_{12}$.
The data were previously published in \citep{Regier_2024}. This material
has space group $R\bar{3}m$, and is antiferromagnetically ordered
below the Néel temperature $T_{\mathrm{N}}\approx8$\,K. The magnetism
arises from the Mn atoms, of which there are 9 per unit cell. Fits
of the magnetic susceptibility above $T_{\mathrm{N}}$ to a Curie-Weiss
law indicated an effective magnetic moment $\mu_{\mathrm{eff}}=1.53\,\mu_{\mathrm{B}}$
per Mn, which is significantly reduced compared to the free-ion value
\citep{Regier_2024}. The measured value of the ordered magnetic moment
$\mu_{\mathrm{ord}}=1.1\,\mu_{\mathrm{B}}$ per Mn from neutron diffraction
\citep{Regier_2024}. The neutron-diffraction pattern of a 4.2-g sample
was measured at $T\approx4$\,K using HB-2A. The sample was loaded
in a 6-mm diameter V container in a closed-cycle refrigerator, the
measurement time for the scan shown was 1.87\,h, and the instrument
used an incident neutron wavelength of $2.41$\,$\text{Å}$ with
pre-sample collimator out \citep{Regier_2024}. 

Fig.~\ref{fig:bcm_sim}(a) compares the experimentally-measured HB-2A
data with the simulated pattern using the same parameters. Excellent
agreement is obtained between the simulation and the measured data.
Fig.~\ref{fig:bcm_sim}(b) shows the different contributions to the
apparent background level in the experiment. In this case, the largest
background contribution is due to incoherent scattering from the V
container, with instrumental background also becoming significant
for $2\theta\lesssim40^{\circ}$. We also note that, if data are collected
or simulated in the paramagnetic phase, there is also a contribution
to the apparent background from paramagnetic diffuse scattering \citep{Paddison_2013}.

Close inspection of the diffraction pattern at $2\theta\lesssim80^{\circ}$
shows some weak peaks that are not accounted for by the nuclear structure
model. These peaks arise from the antiferromagnetic ordering \citep{Regier_2024};
the positions of some magnetic peaks are shown by black arrows in Fig.~\ref{fig:bcm_sim}(c).
The simulated intensities for a ferromagnetic structure are also shown,
assuming 9 magnetic atom per unit cell and magnetic moment magnitudes
of $1.1\,\mu_{\mathrm{B}}$ per magnetic atom. Since the simulated
magnetic structure is ferromagnetic whereas the actual magnetic structure
is antiferromagnetic, the positions of the simulated peaks do not
match the experimental ones. Importantly, however, the magnitude of
the most intense magnetic peak in the simulation is similar to the
most intense magnetic peak in the data. This result suggests that
the simulations can provide a reasonable estimate of the expected
data quality without prior knowledge of the true magnetic structure.

\subsection{Information, warnings, and suggestions}

In addition to plotting the simulated data, the program provides a
list of useful calculated values, including the sample height, its
scattering cross-sections, and its transmission. For absorbing samples,
the optimal sample thickness is approximately $1/\mu$, and the sample
is likely too absorbing if the input container diameter exceeds this
value \citep{Boothroyd_2020}. The program also advises users if the
proposed measurement seems to require adjustment, by providing the
following warnings and suggestions:
\begin{itemize}
\item \emph{Sample height warning}: If the calculated sample height exceeds
6\,cm, the program warns that the sample height exceeds the beam height,
and so the calculation is unrealistic. If the sample height is less
than 3\,cm, the program cautions that the container may not be full.
\item \emph{Low signal-to-background warning}: The program provides this
warning if the ratio of the desired scattering (nuclear coherent or
magnetic) to unwanted scattering (incoherent, multiple, and background)
is less than 0.5. The solution suggested depends on the cause of the
problem: if incoherent scattering is large, isotopic enrichment (e.g.
deuteration) is suggested if possible. If the background scattering
is large and the sample is not too absorbing, increasing the sample
mass is suggested.
\item \emph{Low signal-to-uncertainty warning}: If the ratio of intensity
to uncertainty is less than 30 at the most intense nuclear Bragg peak,
the program suggests increasing counting time, or increasing sample
mass if the sample is not too absorbing.
\end{itemize}

\section{Conclusions and Outlook}

We have developed a web application to enable prospective users of
the neutron-diffraction instruments at the High Flux Isotope Reactor
to predict the quality of data they can expect, given basic information
about their sample and proposed experiment. We hope that this will
be useful to several groups of people. Most importantly, it allows
potential users at HFIR to plan their experiment more effectively.
Second, it allows instrument scientists at HFIR to predict with high
accuracy whether a proposed experiment is likely to provide meaningful
data. Third, we hope this ``recipe'' for simulating powder-diffraction
instruments may help scientists at other neutron-scattering facilities
to develop similar software. The availability of simulation tools
for a wider range of diffractometers would facilitate the identification
of gaps in current neutron-diffraction capabilities worldwide. 

There are several opportunities to extend this program. The simulator
will be updated to reflect upgrades to the instrument, such as the
detector upgrade planned for HB-2A. While we have considered only
constant-wavelength diffraction, extending the methodology to time-of-flight
diffraction measurements would be relatively straightforward. This
would also enable simulations for instruments that are commonly used
for pair-distribution function (PDF) studies, which the effects of
background and counting statistics may be especially important \citep{Olds_2018}.
Similar planning tools could also be developed for spectroscopic data.
We anticipate such simulations may be especially useful for applications
such as phonon dynamics of molecules \citep{Han_2025} and crystal-field
excitations in magnetic materials \citep{Babkevich_2015}. Another
potential application of this program would be to generate large quantities
of realistic simulated data, which are often required for training
machine-learning models \citep{Chen_2021}.

\section*{Acknowledgments}
\begin{acknowledgments}
We are grateful to Malcolm Guthrie (ORNL) and Matt Tucker (ORNL) for useful discussions, and
to the authors of Ref.~\citep{Regier_2024} for allowing re-use of
their published data. This research was supported by the U.S. Department
of Energy, Office of Basic Energy Sciences, Scientific User Facilities
Division. It used resources of the High Flux Isotope Reactor (proposal numbers IPTS-32018 and IPTS-22745) and the
ORNL Research Cloud Infrastructure at the Oak Ridge National Laboratory,
which are supported by the Office of Science of the U.S. Department
of Energy under Contract No. DE-AC05-00OR22725.
\end{acknowledgments}


\begin{thebibliography}{10}

\bibitem{Boothroyd_2020}
A.~T. Boothroyd, {\it Principles of Neutron Scattering from Condensed Matter
  Principles of Neutron Scattering from Condensed Matter\/} (Oxford University
  Press, Oxford, 2020).

\bibitem{Squires_1978}
G.~L. Squires, {\it Introduction to the Theory of Thermal Neutron Scattering\/}
  (Cambridge University Press, Cambridge, 1978).

\bibitem{Lefmann_1999}
K.~Lefmann, K.~Nielsen, {\it Neutron News\/} {\bf 10}, 20 (1999).

\bibitem{Willendrup_2020}
P.~K. Willendrup, K.~Lefmann, {\it Journal of Neutron Research\/} {\bf 22}, 1
  (2020).

\bibitem{Arnold_2014}
O.~Arnold, J.~Bilheux, J.~Borreguero, A.~Buts, S.~Campbell, L.~Chapon,
  M.~Doucet, N.~Draper, R.~{Ferraz Leal}, M.~Gigg, V.~Lynch, A.~Markvardsen,
  D.~Mikkelson, R.~Mikkelson, R.~Miller, K.~Palmen, P.~Parker, G.~Passos,
  T.~Perring, P.~Peterson, S.~Ren, M.~Reuter, A.~Savici, J.~Taylor, R.~Taylor,
  R.~Tolchenov, W.~Zhou, J.~Zikovsky, {\it Nuclear Instruments and Methods in
  Physics Research Section A: Accelerators, Spectrometers, Detectors and
  Associated Equipment\/} {\bf 764}, 156 (2014).

\bibitem{Rodriguez-Carvajal_1993a}
J.~Rodr{\'\i}guez-Carvajal, {\it Physica B\/} {\bf 192}, 55 (1993).

\bibitem{Toby_2013}
B.~H. Toby, R.~B. Von~Dreele, {\it J. Appl. Crystallogr.\/} {\bf 46}, 544
  (2013).

\bibitem{Soper_2011}
A.~K. Soper, {\it GudrunN and GudrunX: programs for correcting raw neutron and
  X-ray diffraction data to differential scattering cross section\/} (Science
  \& Technology Facilities Council Swindon, UK, 2011).

\bibitem{Regier_2024}
C.~E. Regier, S.~O'Donnell, A.~Goyal, M.~J. Dzara, J.~E. Park, R.~T. Bell,
  M.~J. Kramer, J.~A.~M. Paddison, S.~Shulda, D.~S. Ginley, D.~R. Yahne,
  S.~Lany, R.~W. Smaha, R.~A. Klein, {\it Inorganic Chemistry\/} {\bf 63},
  24176 (2024).

\bibitem{Calder_2018}
S.~Calder, K.~An, R.~Boehler, C.~R. Dela~Cruz, M.~D. Frontzek, M.~Guthrie,
  B.~Haberl, A.~Huq, S.~A.~J. Kimber, J.~Liu, J.~J. Molaison, J.~Neuefeind,
  K.~Page, A.~M. dos Santos, K.~M. Taddei, C.~Tulk, M.~G. Tucker, {\it Review
  of Scientific Instruments\/} {\bf 89}, 092701 (2018).

\bibitem{Garlea_2010}
V.~O. Garlea, B.~C. Chakoumakos, S.~A. Moore, G.~B. Taylor, T.~Chae, R.~G.
  Maples, R.~A. Riedel, G.~W. Lynn, D.~L. Selby, {\it Applied Physics A\/} {\bf
  99}, 531 (2010).

\bibitem{Frontzek_2018}
M.~Frontzek, K.~Andrews, A.~Jones, B.~Chakoumakos, J.~Fernandez-Baca, {\it
  Physica B: Condensed Matter\/} {\bf 551}, 464 (2018). The 11th International
  Conference on Neutron Scattering (ICNS 2017).

\bibitem{Frontzek_2018a}
M.~D. Frontzek, R.~Whitfield, K.~M. Andrews, A.~B. Jones, M.~Bobrek,
  K.~Vodopivec, B.~C. Chakoumakos, J.~A. Fernandez-Baca, {\it Review of
  Scientific Instruments\/} {\bf 89}, 092801 (2018).

\bibitem{Lin_2016}
J.~Y. Lin, H.~L. Smith, G.~E. Granroth, D.~L. Abernathy, M.~D. Lumsden,
  B.~Winn, A.~A. Aczel, M.~Aivazis, B.~Fultz, {\it Nuclear Instruments and
  Methods in Physics Research Section A: Accelerators, Spectrometers, Detectors
  and Associated Equipment\/} {\bf 810}, 86 (2016).

\bibitem{Rodriguez-Carvajal_2001}
J.~Rodr{\'\i}guez-Carvajal, {\it FullProf Manual\/}, Laboratoire L{\'e}on
  Brillouin, CEA/Saclay, 91191 Gif sur Yvette Cedex, France (2001).

\bibitem{Hester_2006}
J.~R. Hester, {\it Journal of Applied Crystallography\/} {\bf 39}, 621 (2006).

\bibitem{Momma_2011}
K.~Momma, F.~Izumi, {\it Journal of Applied Crystallography\/} {\bf 44}, 1272
  (2011).

\bibitem{NN_1992}
V.~F. Sears, {\it Neutron News\/} {\bf 3}, 26 (1992).

\bibitem{Hewat_1979}
A.~W. Hewat, {\it Acta Crystallographica Section A\/} {\bf 35}, 248 (1979).

\bibitem{Sabine_1998}
T.~M. Sabine, B.~A. Hunter, W.~R. Sabine, C.~J. Ball, {\it Journal of Applied
  Crystallography\/} {\bf 31}, 47 (1998).

\bibitem{Paddison_2025}
J.~A.~M. Paddison, {\it Journal of Applied Crystallography\/} {\bf 58}, 1022
  (2025).

\bibitem{Blech_1965}
I.~A. Blech, B.~L. Averbach, {\it Phys. Rev.\/} {\bf 137}, A1113 (1965).

\bibitem{Paddison_2013}
J.~A.~M. Paddison, J.~R. Stewart, A.~L. Goodwin, {\it J. Phys.: Condens.
  Matter\/} {\bf 25}, 454220 (2013).

\bibitem{Olds_2018}
D.~Olds, C.~N. Saunders, M.~Peters, T.~Proffen, J.~Neuefeind, K.~Page, {\it
  Acta Crystallographica Section A\/} {\bf 74}, 293 (2018).

\bibitem{Han_2025}
B.~Han, R.~Okabe, A.~Chotrattanapituk, M.~Cheng, M.~Li, Y.~Cheng, {\it Digital
  Discovery\/} {\bf 4}, 584 (2025).

\bibitem{Babkevich_2015}
P.~Babkevich, A.~Finco, M.~Jeong, B.~Dalla~Piazza, I.~Kovacevic, G.~Klughertz,
  K.~W. Kr\"amer, C.~Kraemer, D.~T. Adroja, E.~Goremychkin, T.~Unruh,
  T.~Str\"assle, A.~Di~Lieto, J.~Jensen, H.~M. R\o{}nnow, {\it Phys. Rev. B\/}
  {\bf 92}, 144422 (2015).

\bibitem{Chen_2021}
Z.~Chen, N.~Andrejevic, N.~C. Drucker, T.~Nguyen, R.~P. Xian, T.~Smidt,
  Y.~Wang, R.~Ernstorfer, D.~A. Tennant, M.~Chan, M.~Li, {\it Chemical Physics
  Reviews\/} {\bf 2}, 031301 (2021).

\end{thebibliography}

\end{document}